# Babel Routing Protocol for OMNeT++

More than just a new simulation module for INET framework

Vladimír Veselý, Vít Rek, Ondřej Ryšavý
Department of Information Systems, Faculty of Information Technology
Brno University of Technology
Brno, Czech Republic
{ivesely, rysavy}@fit.vutbr.cz; rek@kn.vutbr.cz

*Abstract*—Routing and switching capabilities of computer networks seem as the closed environment containing a limited set of deployed protocols, which nobody dares to change. The majority of wired network designs are stuck with OSPF (guaranteeing dynamic routing exchange on network layer) and RSTP (securing loop-free data-link layer topology). Recently, more use-case specific routing protocols, such as Babel, have appeared. These technologies claim to have better characteristic than current industry standards. Babel is a fresh contribution to the family of distance-vector routing protocols, which is gaining its momentum for small double-stack (IPv6 and IPv4) networks. This paper briefly describes Babel behavior and provides details on its implementation in OMNeT++ discrete event simulator.

I. INTRODUCTION

Currently deployed routing protocols, e.g., OSPF, RIP, EIGRP, have well-known characteristics. New proposals that address shortcomings of standard routing protocols need to be evaluated to reach the required level of maturity to be widely adopted by the industry. Simulation approach can provide an initial evaluation of a new routing protocol and its comparison to existing protocols. Presented paper provides notes on development and evaluation of a simulation model for Babel routing protocol.

The newly developed Babel simulation model is a part of ANSAINET framework[1] which aims at providing a variety of simulation models compatible with RFC specifications and reference implementations. These tools should allow for analysis of a near real network behavior. Results are freely available and can be employed for further research initiatives, such as simulation approach to proving (or disproving) certain aspects of technologies and related protocols. The ANSAINET now supports:

- Babel dynamic unicast routing protocol;
- Proprietary first-hop redundancy protocols (FHRP) such as Hot Standby Router Protocol (HSRP) and Gateway Load Balancing Protocol (GLBP), which guarantee high-availability of default gateway;
- Device discovery protocols such as Cisco Discovery Protocol (CDP) and Link Layer Discovery Protocol (LLDP), which verify data-link layer operation.

In this paper, we only focus on a Babel simulation model. Babel is increasingly more popular seen as the open-source alternative to Cisco's Enhanced Interior Gateway Routing Protocol (EIGRP). Babel is also considered a better routing protocol for mobile networks comparing to Destination-Sequenced Distance-Vector (DSDV) or Ad hoc On-Demand Distance-Vector (AODV) routing protocols. Babel is a *hybrid distance vector* routing protocol. Although it stems from a classical distributed Bellman-Ford algorithm, it also adopts certain features from link-state protocols, such as proactive neighbor discovery. It offers a great modularity of metric calculation (currently four distinct techniques are specified) and pluggable best route selection policy. Babel employs a *feasibility condition test* to prevent route loops during convergence phase. Babel protocol syntax employs type-length-value (TLV) encoding, which for instance allow incorporating different address-family (currently both IPv4 and IPv6) for routing information. Babel has a built-in mechanism for duplicity prevention together with a data compression that reduces protocol overhead.

This paper has the following structure. Section II covers a quick overview of existing implementations. Section III describes the operational theory and design supplemented with implementation notes. Section IV contains validation and testing scenarios. The paper is summarized in Section V together with outlines of our plans.

II. CURRENT BABEL IMPLEMENTATIONS

This section brings brief information about the availability of Babel for real network deployment. According to the main Babel project web page [3], there are three active (and one deprecated) implementations available:

- *babeld* – reference implementation maintained by the author of Babel specification [4];
- *Pybabel* – implementation in Python limited only to IPv6 and no real cost recalculation [5];
- *Sbabeld* – a limited implementation intended only for IPv6 stub-routers [6];

---
[1] ANSA is a long-term project carried by researchers and students at the Brno University of Technology aiming at extending IP network simulation framework with new simulation models. The framework is built on the original INET framework [1] shipped with OMNeT++ simulator.

We have unsuccessfully searched for any Babel simulation model. In particular, we are not aware of any Babel implementation in the most used discrete event simulators, such as NS-2/3, OPNET, or OMNET++.

### III. CONTRIBUTION

The aim of this section is to overview the basic elements of Babel: 1) the best route selection, 2) message exchange mechanisms, and 3) state maintenance structures. Also, we provide an explanation of how these elements are implemented in Babel's simulation model of ANSAINET.

#### A. Theory of Operation

Babel is codified withing IETF as experimental RFC 6126 [7]. It leverages both unicast communication and also multicast address 224.0.0.111 for IPv4 and ff02::1:6 for IPv6 group communication. Babel operates over UDP on (both source and destination) port 6696.

Babel is using **feasibility condition (FC)** when verifying incoming routing records. In particular, Babel employs FC variant called *Source Node Condition* [8] just as EIGRP: The best known metric $m_A$ together with a sequence number $s_A$ (number reflecting age of metric, higher means younger and more current) to a destination network $N$ from a router $A$ denotes its feasible distance, $FD_A(N) = (s_A, m_A)$. Routing information received by router $A$ from router $B$ satisfies FC if and only if the metric $D_B(N)$ to the destination network $N$ advertised by router $B$ is strictly lower than $FD_A(N)$:

$$D_B(N) = (s_B, m_B), FD_A(N) = (s_A, m_A):$$
$$D_B(N) < FD_A(N) \leftrightarrow (s_B = s_A \wedge m_B < m_A) \vee s_B > s_A$$

Applying FC, we avoid counting-to-infinity problem known from original RIP implementation. However, the FC might cause *starvation*, when the only one route exists, and it cannot be used because it does not satisfy FC. Therefore, Babel is checking sequence numbers (just as DSDV) to recognize outdated $FD$. Moreover, Babel equips routing updates also with advertising router identification to distinct between different routes to the same network prefix.

Babel employs the sum of links costs from the router to a given destination network as the metric. Babel evaluates metric on router $A$ as the route's metric $m_B$ announced by a neighbor $B$ plus link's cost $c$ between $A$ and $B$: $m_A = m_B + c$ Babel currently suggests two methods how link's cost may be calculated:

- *k-out-of-j* – This method is intended for wired networks using two parameters $j$ and $k$, where $0 < k \leq j$. A router remembers window of size $j$ containing last $k$ Hello messages. If less than $k$ Hello messages have been delivered, then cost is set to 0xFFFF, which means infinity (unreachable network), otherwise ($k$ and more *Hellos* have been successfully delivered) cost is set to a fixed value reflecting the link's speed (by default 96 on wired and 256 on wireless interfaces);

- *ETX* (based on [9]) – This method targets specifics of wireless networks. Link cost varies in time, and it is determined based on two parameters – successful *Hello* reception ($\beta$) and successful *Hello* transmission ($\alpha$). Aggregated link cost is computed as $256/(\alpha \cdot \beta)$.

Babel exchanges data as TLV records. Babel message consists of several TLVs records, which is controlled by a buffering policy. Babel records are:

- *AckReq* and *Ack* – AckReq is the ack request, and *Ack* is the solicited acknowledgment response within specified interval using the same nonce;

- *Hello* – Neighbor and link's reception cost are discovered using *Hello*;

- *IHU – I Hear You* is confirmation of mutual reachability of neighbors. Moreover, these TLVs carry link's transmission cost;

- *Router-Id* – TLV contains unique (recommended is to use EUI-64) eight bytes long router identification within a given Babel routing domain;

- *NextHop* – TLV carries next-hop IP address for subsequent *Updates* TLV;

- *Update* – It advertises or withdraws routes including their prefix, two bytes long sequence number and metric;

- *RouteReq* – TLV prompts receiver to send *Update* regarding specified prefix;

- *SeqNoReq* – It prompts receiver to send (or to delegate further) *Update* regarding specified prefix with a given sequence number.

- *Pad1* and *PadN* – These two are padding TLVs without any meaningful content;

Babel standard specifies multiple abstract data structures and placeholders for information important for Babel functionality:

- *Interface Table* contains the list of interfaces through which Babel routing information are being sent and accepted;

- *Neighbor Table* (NT) contains the list of all known neighbors including mutual interface (through which neighbor is reachable), IP address, history of *Hellos*, transmission cost, sequence number (used by neighbor);

- *Source Table* (ST) is a placeholder for *FD*s of different network prefixes. Each record is specific for a given advertising router and prefix (including prefix length) and contains $(s, m)$ tuple;

- *Route Table* (RT) contains Babel routes (tuples of prefix, prefix length, next-hop, neighbor's router-id and address, metric, sequence number) known to this router;

- *Pending Requests Table* (PRT) maintains the list of sent but not yet answered requests during network topology convergence time;

#### B. Simulation Model Implementation

In OMNet++, Babel is implemented as the compound module `BabelRouting` interconnected via `UDP` sockets with IPv4 and IPv6 network layers. It consists of three submodules that are depicted in Fig. 1 and briefly described in Table I. Our implementation is in full compliance with standard RFC 6126.

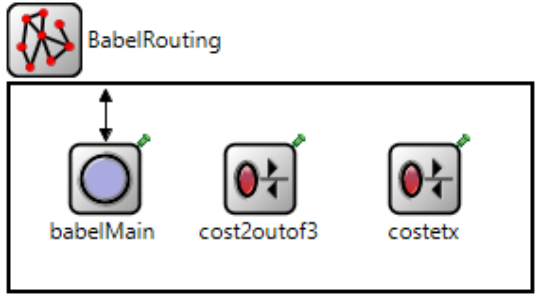

Fig. 1: BabelRouting module structure

TABLE I. DESCRIPTION OF BABEL ROUTING SUBMODULES

| Name | Description |
|---|---|
| babel Main | Implements Babel routing protocol behavior. Module contains auxiliary classes such as: `BabelInterfaceTable` (contains the list of Babel enabled interfaces including relevant settings); `BabelNeighborTable` (maintains the state of neighborship with adjacent routers); `BabelTopologyTable` (all routes towards known destinations); `BabelSourceTable` (table containing *FD*s, network prefixes and identifiers of routers propagating routing information); `BabelPenSRTable` (the list of pending requests); `BabelFtlv` (internal representation of uncompressed TLV records); and `BabelBuffer` (sending buffer for combining multiple TLVs into one outgoing message). |
| cost 2outof3 | Implements link cost calculation employing method *k-out-of-j* with $k = 2$ and $j = 3$. This algorithm is usually used in wired networks. |
| cost etx | Implements link cost calculation using method ETX, which is suitable for wireless networks. |

The user may easily repeat the design pattern of `cost*` modules to create a new policy of link cost calculation.

Babel behavior relies on timers. Following eleven timers are fundamental and corresponds to various conditions:

- *Hello timer* specifies time between two consecutive *Hello* messages (by default 20 seconds on wired and 4 seconds on wireless links);

- After *Update timer* expiration, periodic *Updates* are sent (by default 4× *Hello timer*);

- *Buffer* and *BufferGC timers* are employed to remove (un)used data from buffer;

- *ToAckResend timer* governs retransmission of *Ack*;

- *NeighHello* and *NeighIHY* timers hold neighbor's *Hello* and *IHY* intervals;

- *RouteExpiry* and *RouteBefExpiry* affect each RT's records validity period;

- Each record in ST has its *SourceGC* timer; Babel removes the record from ST after the timer's expiration;

- *SRResend* timer triggers PRT's request resending.

## IV. TESTING

In this section, we provide information regarding validation of our simulation model against existing implementations. The goal is to demonstrate that this module is accurate enough for simulation of real network scenarios.

For test cases, we have built the same *real* network topology as for the simulation. We captured and analyzed (using Wireshark 1.12.7 packet sniffer) relevant Babel messages exchanged between routers. The routers run *babeld* v1.5.1 on operating system Debian 7.7 x64 (kernel 3.2.65-1).

Topology, which is depicted in Fig. 2, contains four routers (R1, R2, R3, and R4), four Local Area Network (LAN) segments (either simulated using dual-stack host or directly connected interfaces) and interconnections between routers. Topology operates both IPv4 and IPv6 address families. Babel *2-out-of-3* strategy is employed for link cost calculation. *Hello interval* is set to 4 seconds, and split-horizon is activated on all interfaces.

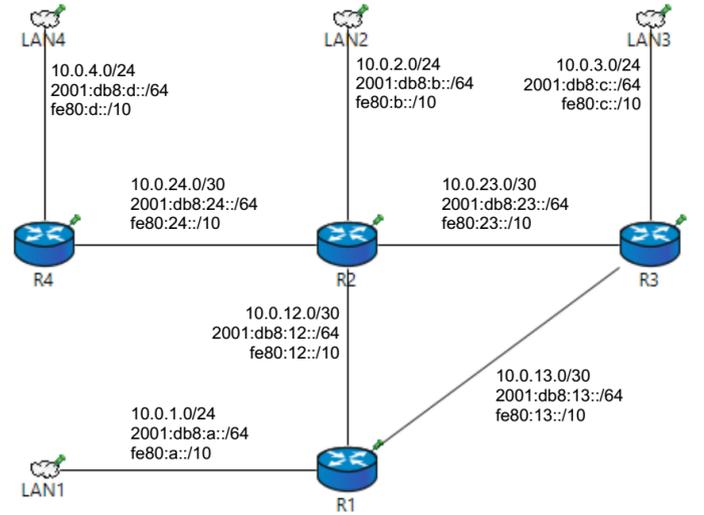

Fig. 2: Babel testing topology

We have conducted three tests to compare the behavior of implementation in common scenarios for Babel protocol: (A) establishing neighborship, (B) routing table convergence, and (C) link failure and subsequent routing information propagation. The details and results are described in the rest of this section.

### A. Establishing Neighborship

This test aims at observing initial Babel message exchange between two adjacent routers when they discover each other on the link. We have decided to describe in more detail situation between routers R1 and R2 because establishing neighborship among any other two adjacent routers is analogous. We align events between simulation and real network by the time $t_0$ when router R1 initiates Babel process while R2 is already operational. We achieve this alignment: a) by executing *babeld* process in the real network; b) by connecting link between routers in the simulation. Table II summarizes exchanged TLVs. Column marked with "Ord." shows the order of intercepted messages. Column marked with header "S → R" contains sender and receiver of a given message. Timestamps in "Simul." and "Real" columns are relative to the $t_0$ event.

TABLE II. ESTABLISHMENT NEIGHBORHOOD TIMESTAMP COMPARISON

| Ord. | TLVs | S → R | Simul.[s] | Real [s] |
|---|---|---|---|---|
| #1 | Hello, RouteReq | R1→R2 | 0.092 | 0.006 |
| #2 | Hello, IHU, Update | R2→R1 | 0.292 | 0.007 |
| #3 | Hello, IHU | R1→R2 | 0.492 | 0.040 |
| #4 | Hello, IHU | R2→R1 | 0.692 | 0.134 |
| #5 | RouteReq | R2→R1 | 0.692 | 0.903 |
| #6 | Hello, IHU, *Update* | R1→R2 | 0.892 | 1.084 |
| #7 | RouteReq | R1→R2 | 0.892 | 1.085 |
| #8 | Update, *IHU* | R2→R1 | 1.902 | 1.744 |
| #9 | Hello, IHU | R2→R1 | 5.632 | 5.111 |

Babel message #1 (containing *Hello* and *RouteReq* TLVs) announces R1 to be present on the link. R2 responds with #2 appending known routes in *Update* TLV. It takes #3 and #4 messages to pronounce neighbors as operational because of *2-out-of-3* strategy. R2 sends #5 directly to R1 via unicast asking R1 for known routes. R1 appends *Update* as the reaction to the previous message. R1 repeats the similar process of route discovery via *RouteReq* in #7, where R2 responds with #8. Expired *Hello timer* causes R2 to send message #9.

Note that `babeld` can cope with lossy links by sending several copies of the same TLV (marked red in Table II). This advanced behavior not described in RFC and thus not implemented by the current version of the Babel simulation model.

### B. Routing Table Convergence

During this test, we analyze network convergence process and compare the content of RTs in the resulting stable state. We focus on the IPv4 routes content in RT. Fig. 3 shows its content in the simulation and Table III in the real network. We can find the slight difference (marked with red) between simulated and real network. Destination network 2001:db8:23::/64 is reachable from R1 via two routes – the first goes through R2 and the second through R3. They have the same metric, and both of them are equally reachable. As Babel does not support load-balancing only one route can be selected. The selection of the route is implementation dependent. It explains why the simulation experiment differs to the real network experiment in this partial result.

TABLE III. R1'S ROUTE TABLE CONTENT IN THE REAL NETWORK

| Flag Prefix | Met | RD | Router-Id | Next-Hop |
|---|---|---|---|---|
| > 2001:db8:a::/64 | 0 | | | |
| > 2001:db8:12::/64 | 0 | | | |
| > 2001:db8:13::/64 | 0 | | | |
| > 2001:db8:b::/64 | 96 | 0 | 2222:2222:2222:2222 | fe80:12::2 |
| > 2001:db8:c::/64 | 96 | 0 | 3333:3333:3333:3333 | fe80:13::3 |
| > 2001:db8:d::/64 | 192 | 96 | 4444:4444:4444:4444 | fe80:12::2 |
| 2001:db8:12::/64 | 96 | 0 | 2222:2222:2222:2222 | fe80:12::2 |
| 2001:db8:13::/64 | 192 | 96 | 3333:3333:3333:3333 | fe80:12::2 |
| 2001:db8:12::/64 | 192 | 96 | 2222:2222:2222:2222 | fe80:13::3 |
| 2001:db8:13::/64 | 96 | 0 | 3333:3333:3333:3333 | fe80:13::3 |
| > 2001:db8:23::/64 | 96 | 0 | 2222:2222:2222:2222 | fe80:13::2 |
| 2001:db8:23::/64 | 96 | 0 | 3333:3333:3333:3333 | fe80:13::3 |
| > 2001:db8:24::/64 | 96 | 0 | 2222:2222:2222:2222 | fe80:12::2 |

Fig. 3. R1's Route Table content in the simulation

### C. Link-failure

We have scheduled link failure (physically disconnecting the link) between R1 and R2 at $t_0$. We analyze the impact on reachability of network 2001:db8:a::/64 from the perspective of router R2. Fig. 4 shows RT:

- before the failure (the next-hop is R1 with address fe80:12::1 and metric 96);
- shortly after the failure (with poisoned metric 65535);
- after a while, when network correctly converges to R3 as the new next-hop (with address fe80:23::3).

Fig. 4. R2's route state before/after the link failure and after convergence

This scenario demonstrates the usage of sequence numbers by Babel. If a backup route satisfying FC was available, then it would be immediately used. However, the poisoned route cannot be removed from the RT and updates from R3 are ignored. Babel solves this starvation situation by incrementing route's sequence number. Related communication is outlined in Table IV, where timestamps are aligned with $t_0$ failure event.

TABLE IV. LINK FAILURE TIMESTAMP COMPARISON

| Ord. | TLVs | S → R | Simul.[s] | Real [s] |
|---|---|---|---|---|
| #1 | SeqNoReq | R2→R3 | 0.187 | 0.208 |
| #2 | SeqNoReq | R3→R1 | 0.347 | 1.079 |
| #3 | Update | R1→R3 | 0.595 | 1.152 |
| #4 | Update | R3→R2 | 0.673 | 1.275 |

Table IV reveals noticeable timestamp differences between simulated and real network environments in this test. Nevertheless, the order of messages is still preserved. Time variations are caused by three factors: a) *babeld* operation is influenced by operating system interrupts; b) built-in packet pacing avoiding potential race conditions; and c) inaccuracy in timing of $t_0$ event in the real network. Nevertheless, the routing tables in the simulated and real networks are equivalent.

## V. Conclusion

In this paper, we presented a description of Babel dynamic routing protocol and its implementation as a simulation model for OMNeT++. We have designed and implemented a new simulation model mimicking the behavior of full-fledged Babel implementation for OMNeT++. Moreover, we have verified the accuracy of simulation results taking into account real wired network baselines.

Developed Babel simulation model is also a tool for the next steps of our future work. We plan to compare the properties of various routing protocols in specific scenarios (e.g., high availability data-centers). Another possibility would be to employ Babel in wireless scenarios but it is not our primary goal since ANSAINET is focused on traditional wired networks and their technologies. We hope that our contribution will be eventually included in the official INET release. Moreover, Babel module is also going to be employed in the frame of PRISTINE project [10] as a support of ongoing research.

All source codes could be downloaded from ANSAINET GitHub repository [11]. Real packet captures and simulation datasets for results reproduction could be obtained from Wiki pages of the above-mentioned repository. More information about ANSA project is available on its homepage [12].

## Acknowledgment

This work was supported by the Brno University of Technology organization and by the research grants:

- FP7-PRISTINE supported by the European Union;
- FIT-S-14-2299 supported by Brno University of Technology;